\documentclass[reprint,aps,prl,twocolumn,superscriptaddress]{revtex4-1}
\usepackage{braket}
\usepackage{amssymb}
\usepackage{amsmath}
\usepackage{epsfig}
\usepackage{color}
\usepackage{graphics, graphicx}
\usepackage{bbold}
\usepackage{psfrag}
\usepackage{mathcomp}
\usepackage{subfigure}
\usepackage{verbatim}
\usepackage[colorlinks, citecolor=blue]{hyperref}
\usepackage[normalem]{ulem}
\usepackage{bm}
\begin{document}

\title{Comment on "Shell-Shaped Quantum Droplet in a Three-Component Ultracold Bose Gas"}

\maketitle

In the Letter [1], Ma et al. propose a new type of shell-shaped
Bose-Einstein condensate with a self-bound character, made of three-component 
$Na^{23}K^{39}K^{41}$ Bose mixture (species (1,2,3) in the following),
where the mixtures (1, 2) and (2, 3) both form quantum droplets. 
The proposed structures are made of an outer shell of liquid (1,2) enveloping a spherical 
core of (2,3) liquid, which is claimed to be stable without the need of any trapping potential.
I comment in the following that these structures are not actually the ground-state
solutions to the system but rather local energy minima, and most likely impossible to 
realize in practice.
The lowest energy structure is instead (for all the cases considered in [1])
a "dimer" configuration, composed
of two droplets (made of the 1-2 fluid and 2-3 fluid, respectively)
kept in mutual contact by the shared component 2.
I show in Fig.1 one of the shell structures discussed 
in Ref.[1], obtained here using their
same theoretical method (upper panel), 
and the ground-state "dimer" for the same mixture (lower panel).
\begin{figure}[t]
\includegraphics[width=7cm]{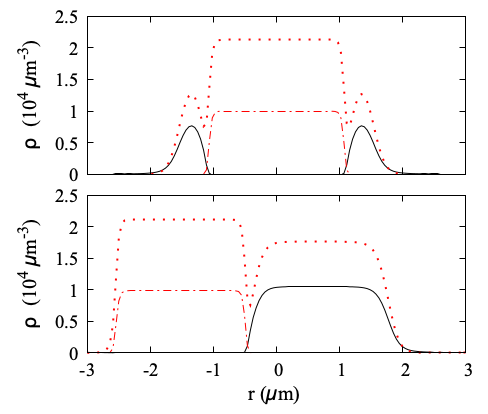}
\caption{Density profile along the line passing through the center of the droplets.
Panel a): shell-structure with liquid (1,2) enveloping the (2,3) core;
lower panel: ground-state for the same system ("dimer" structure).
Solid line: species 1;dotted line: species 2; dash-dot line: species 3.
} 
\label{fig1}
\end{figure}
For clarity, I show in Fig.2 a map of the total density of the system in
the central plane for the two configurations. 
It appears that the fluid (1,2) (droplet on the right), rather than "wetting" 
uniformly the inner spherical core of liquid (2,3), as in 
shell-shaped configurations, prefers a non-wetting
configurations where it minimizes the contact with the (2,3) fluid 
surface (droplet on the left).
This non-wetting behavior is stronger the thinner is the outer shell: indeed,
of the three different shell widths investigated in Ref.[1] for the same
value of the scattering length $a_{23}$, I find that the less stable
is the one with the thinnest outer shell. In this case, once this 
configuration is reached by carefully choosing
the initial density profiles, even a very small amount of random noise can
destabilize it, eventually ending, during the energy minimization, 
into the more stable "dimer" state.
\begin{figure}[t]
\includegraphics[width=7cm]{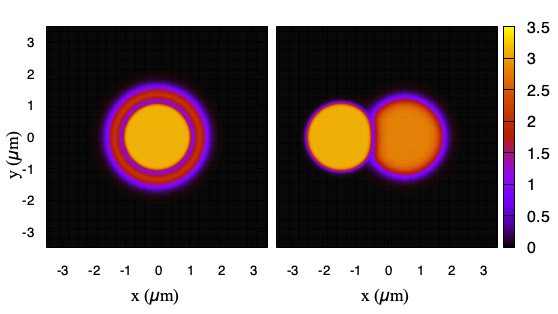}
\caption{Total density maps corresponding to the two
structures in Fig.1. Left: shell-shaped structure; Right: ground-state "dimer" solution.
} 
\label{fig2}
\end{figure}
In order to obtain the shell-shaped structures 
like the one shown in the upper panel of Fig.1,
the energy minimization must start from a
carefully chosen initial guess for the density.
In Ref.[1] they suggest that a possible way to realize in practice 
a shell-shaped configuration
would consist in 
preparing the core-shell structure in an isotropic
harmonic trap, and then release it.
However, this does not provide the desired configuration: 
I find that the ground-state of a uniform mixture initially confined
in an isotropic harmonic trap 
assumes indeed a shell structure during the energy minimization, 
but where the outer shell is made by (2,3) liquid
and the inner core by the (1,2) liquid. Once the trap is released, however, 
this shell structure
swiftly relaxes towards a lower energy state represented 
by a spherical core of liquid (2,3) with six equal droplets of liquid (1,2)
symmetrically attached to it in the geometry of a regular octahedron,
again showing the tendency of the (1,2) liquid to avoid wetting the (2,3) fluid 
but rather minimize the interface area between the two liquids.
In the light of the above findings, the claim made in Ref.[1] that 
quasi 2D shells with extremely thin width can be realized in equilibrium with 
a spherical core of large radius also appears unrealistic.

\smallskip
Work supported by the Italian MIUR under the PRIN2022 Project No. 20227JNCWW.

\medskip
\noindent
\begin{large}Francesco Ancilotto\newline \noindent
\end{large} 
Dipartimento di Fisica e Astronomia ``Galileo Galilei''\newline \noindent 
Universit\`a di Padova, via Marzolo 8 - Padova, Italy;\newline \noindent
CNR-IOM, via Bonomea, 265 - Trieste, Italy 

\medskip
[1] Y. Ma and X. Cui, Phys. Rev. Lett. {\bf 134}, 043402 (2025).
\end{document}